\documentclass[conference]{IEEEtran}
\IEEEoverridecommandlockouts

\usepackage{todonotes}
\usepackage{cite}
\usepackage{amsmath,amssymb,amsfonts}
\usepackage{algorithmic}
\usepackage{graphicx}
\usepackage{textcomp}
\usepackage{xcolor}
\def\BibTeX{{\rm B\kern-.05em{\sc i\kern-.025em b}\kern-.08em
    T\kern-.1667em\lower.7ex\hbox{E}\kern-.125emX}}
\begin{document}

\title{Effects of Limited Field of View on Musical Collaboration Experience with Avatars in Extended Reality}

\author{ 

\IEEEauthorblockN{Suibi Che-Chuan Weng}
\IEEEauthorblockA{\textit{ATLAS Institute} \\
\textit{University of Colorado, Boulder}\\
Boulder, USA \\
suibi.weng@colorado.edu}
\and

\IEEEauthorblockN{Torin Hopkins}
\IEEEauthorblockA{\textit{SolJAMM Research}\\
Denver, USA \\
0000-0001-7359-2906}
\and
\IEEEauthorblockN{Shih-Yu Ma}
\IEEEauthorblockA{\textit{Department of Computer Science}\\
\textit{University of Colorado Boulder} \\
Boulder, USA \\
shih-yu.ma@colorado.edu}
\and

\IEEEauthorblockN{Amy Bani\'{c}}
\IEEEauthorblockA{
\textit{Interactive Realities Lab} \\
\textit{Department of Electrical Engineering and Computer Science} \\
\textit{University of Wyoming}\\
Laramie, USA \\
abanic@uwyo.edu}
\and
\IEEEauthorblockN{Ellen Yi-Luen Do}
\IEEEauthorblockA{\textit{ATLAS Institute} \\
\textit{University of Colorado, Boulder}\\
Boulder, USA \\
ellen.do@colorado.edu}

}

\maketitle


\IEEEpeerreviewmaketitle

\begin{abstract}
During musical collaboration, visual cues are essential for communication between musicians. Extended Reality (XR) applications, often used with head-mounted displays like Augmented Reality (AR) glasses, can limit the field of view (FOV) of players. We conducted a study to investigate the effects of limited FOV on co-presence, gesture recognition, overall enjoyment, and reaction time.

Initially, we observed experienced musicians collaborating informally with and without visual occlusion, noting that collaboration suffered with limited FOV. We then conducted a within-subjects study with 19 participants, comparing an unrestricted FOV holographic setup called HoloJam to Nreal AR glasses with a 52$^{\circ}$ limited FOV. In the AR setup, we tested two conditions: standard AR with a 52$^{\circ}$ FOV and a modified AR notification system called Mini Musicians.

Results showed that HoloJam provided higher co-presence, quicker gesture recognition, and greater enjoyment. The Mini Musicians application reduced reaction time and maintained enjoyment compared to the standard AR setup. We conclude that limited FOV impacts musical collaboration, but notifications can improve reaction time and should be considered in future XR music collaborations.

\end{abstract}



\section{Introduction}

\begin{figure*}[!t]
  \centering
  \includegraphics[width=\linewidth]{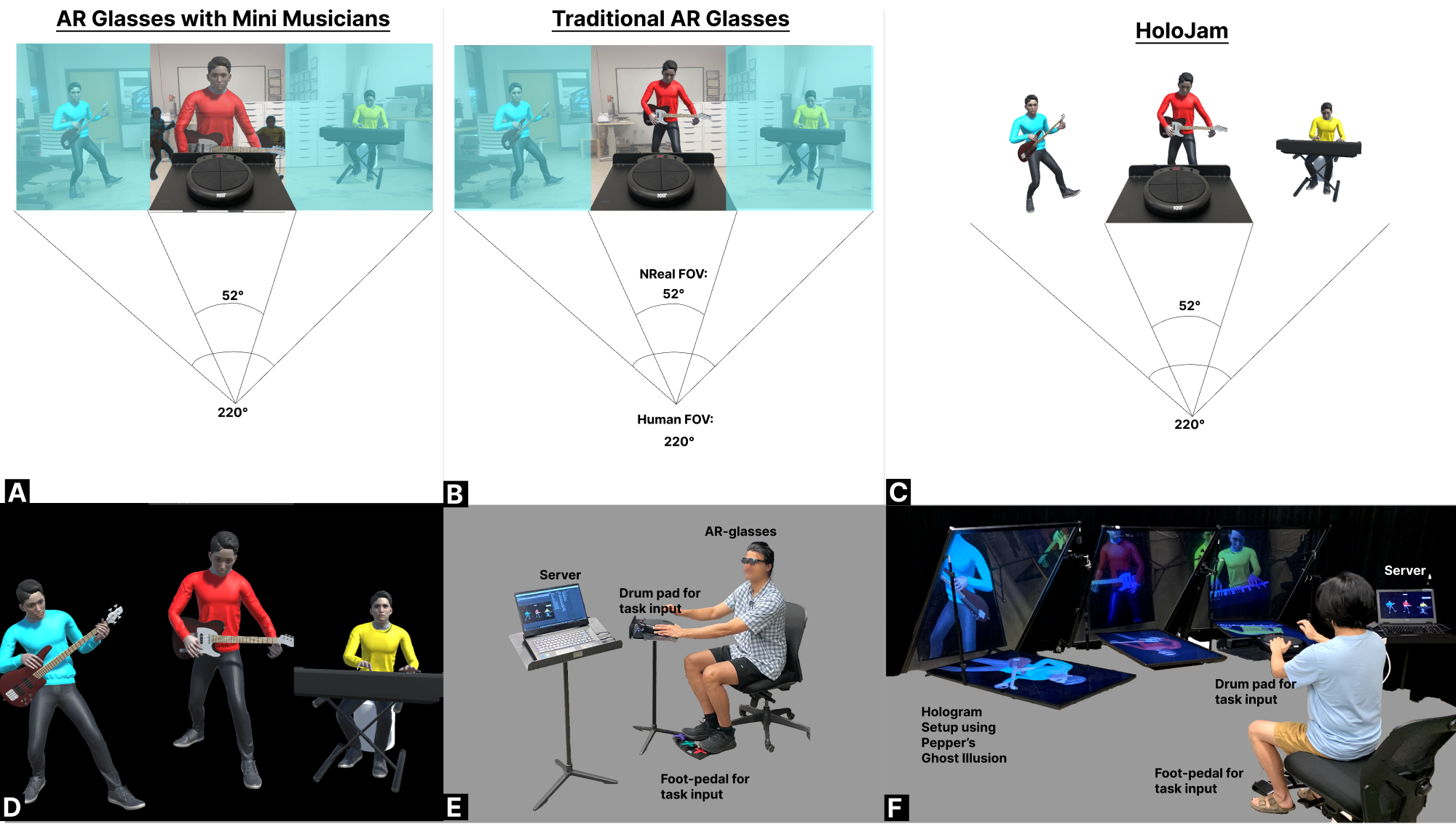}
  \caption{A) A representation of the experimental setup using AR glasses, a MIDI drum, and mini musicians; 
  B) An illustration of limited field of view with no mini musicians; 
  C) Demonstrates the holographic setup with no limit on field of view.
  }
  \label{fig:teaser}
\end{figure*}

During the COVID-19 pandemic, individuals continued to seek ways to connect with others through music, including engaging in networked music platforms \cite{Hopkins22,jamulus,jacktrip}. Remote music collaboration software such as Jacktrip \cite{jacktrip} and Jamulus \cite{jamulus} facilitated playing music together at a distance and aimed to recreate an authentic auditory musical experience. While networked video facilitates visual collaboration \cite{jacktrip}, the use of Extended Reality (XR) applications presents an alternative avenue to explore social dynamics in three dimensions\cite{Hopkins_XRJAM_2024,Boem_MetaMusic_2024}. In musical collaboration, communication between players occurs spatially, usually involving directed hand gestures, facial expressions, and eye-gaze \cite{Hopkins22}. This information is essential for musical transitions and collective musical coordination.

Recently, the use of AR headsets has seen a substantial increase, sparking a heightened interest in the study of interactions between humans and virtual content in augmented reality. As AR technology progresses, understanding how people perceive and interact with virtual elements becomes increasingly important, despite current limitations like a restricted field of view (FOV). Research in this area often examines how users process information that lies outside of their FOV \cite{trepkowski_effect_2019,piumsomboon_mini-me_2018}. Much of this research focuses on enhancing task performance through message notifications rather than improving the social presence of avatars outside the FOV \cite{gruenefeld_flyingarrow_2018,gruenefeld_radiallight_2018,qian_restoring_2018,fan_spidervision_2014}. Musical collaboration, which heavily depends on communicative gestures, provides a valuable context for studying interactions with avatars \cite{pai_neuraldrum_2020,Hopkins22}. While gestures like hand movements and eye contact are easily observed in face-to-face interactions, detecting these social cues becomes challenging in multi-avatar collaborations when avatars are outside the user's FOV.


In our study, we provided a more detailed investigation building upon the research by Weng et al. \cite{Weng_HolojamISMAR_2023,Weng_HolojamVR_2023} regarding the impact of FOV limitations on musical collaboration.

\begin{figure*}[!tb]
 \centering 
 \includegraphics[width=1\linewidth]{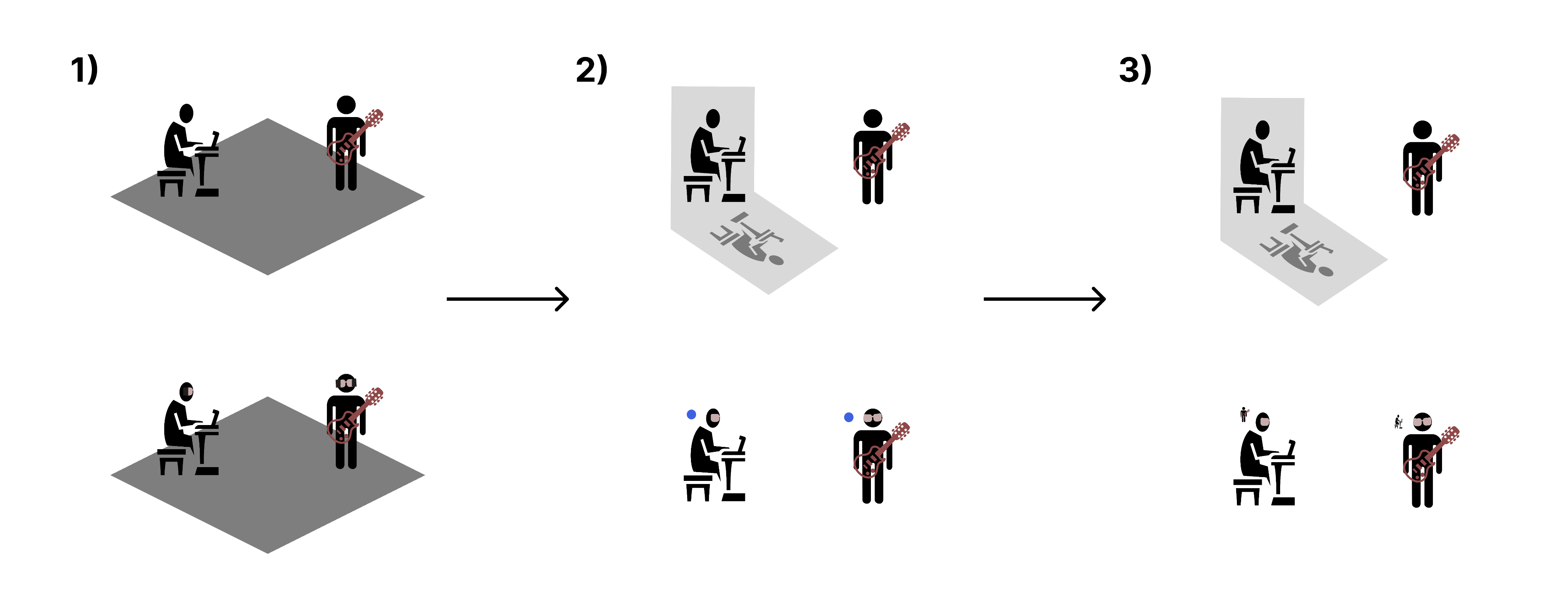}
 \caption{(1) Assessing in person jam sessions with and without visual occlusion at 52 degrees (2) Assessment of holograms vs blue light notification,~\cite{Weng_HolojamISMAR_2023} (3) Assessment of holograms vs mini musicians~\cite{Weng_HolojamVR_2023}}
 \label{fig:studies}
\end{figure*}

To delve deeper into this problem, we carried out an investigation in three parts, shown in Figure \ref{fig:studies}. 
Insights from these sessions and previous work on AR notifications led researchers to test a holographic musician system against a known type of peripheral view notification—blue spheres.\cite{gruenefeld_radiallight_2018,Weng_HolojamISMAR_2023}.Researcher noticed that reaction times were shorter when participants used the notification system, although they reported feeling distracted. \cite{Weng_HolojamISMAR_2023} 

Additionally,we observed live jam sessions where musicians played music together under two conditions: 1) normally, and 2) while wearing glasses that blocked their peripheral vision.Building on our observations and previous related work, we developed three systems to test our hypothesis. The first system features Holojam which was a three holograms created using the Pepper's ghost illusion, representing a condition without any FOV limitations (Figure \ref{fig:teaser} C).  The second condition uses the exact same application but uses AR glasses (Figure \ref{fig:teaser} A and B) and user's position, with a limited FOV of 52$^\circ$. The third and final condition is the same application, but when avatars are outside the field of view a small and semi-transparent version of the avatar appears within the field of view (mini musicians). The notification system was called Mini Musicians.We conducted a within-subject study comparing the three conditions. In the study participants were asked to play a hand drum with three avatar musicians and indicate when the avatar is looking at them.

Throughout this paper we discuss work that inspired our experimental design, methods and results, and implications for future work in the field. Results showed that a limited FOV does affect musical collaboration experience. Conditions associated with a larger field of view yielded a significantly higher score in sense of co-presence with the avatars and overall enjoyment. However, the notification strategy of mini musicians reduced the reaction time significantly with AR glasses conditions.

\textbf{Contributions of our study are:}

\begin{itemize}
\item A limited FOV significantly reduces musicians ability to communicate during a jam session, which impacts the music and perceived experience
\item The mini musicians notification helps to reduce the reaction time with AR glasses
\item Notifications while playing music in AR do not significantly affect co-presence, likely because of distraction
\end{itemize}

\section{Related Work}
Our research focuses on three distinct areas. First, we delve into the realm of musical collaboration in extended reality (XR), examining how avatars can enhance the immersive experience for musicians engaged in cooperative musical activities. Second, we explore the development of notifications specifically designed for AR glasses, aiming to address the challenge of objects that are outside the user's field of view. By providing visual cues or indicators for these objects, we aim to enhance the overall AR experience and improve user engagement. Lastly, we investigate the concept of social presence in avatars within the AR context, examining how users perceive and interact with avatars in virtual environments, with a particular emphasis on their impact on cooperative interactions and perceived presence.

\subsection{Musical collaborations in XR }

Over the past two decades, the intersection of music and XR has become a well-established area of research, transforming traditional musical interactions by enabling players to engage with virtual objects, agents, and environments \cite{Turchet2021}. During the COVID-19 pandemic, face-to-face music collaboration was limited, leading to a burgeoning exploration of mixed reality technologies as a means to create a sense of presence and community. The project Wish You Were Here ~\cite{Schlagowski2023}, is a laboratory study conducted to assess the potential of mixed reality for music collaboration, comparing it to an audio-only baseline. The results showed increased co-presence and positive affect for musicians jamming in mixed reality. 

Various other methods have been explored for collaborative music-making, such as using a composer table~\cite{Berry2003,Jord2007,smartcities} or employing a graphic interface to document the collaborative experience~\cite{Bryan-Kinns2004,Men19}. However, the use of AR or virtual reality (VR) has demonstrated superior immersive experiences in musical collaborations ~\cite{pai_neuraldrum_2020,cairns2023evaluation}. Studies, including the AR drum circle~\cite{Hopkins22}, have emphasized the significance of social cues in XR musical collaborations, where animated avatars enhance the immersive experience. Nevertheless, challenges remain, such as the limited FOV of AR glasses, which can lead to virtual characters being cut off, affecting communication and musical performance. Our study aims to provide statistical insights into the impact of restricted FOV on musical collaborations with avatars via AR glasses.

\subsection{Notifications for limited FOV of AR glasses}
Many optical–see-through displays have a relatively narrow ﬁeld of view. However, a limited FOV can constrain how information can be presented and searched through ~\cite{trepkowski_effect_2019}. Researchers have proposed solutions at both the hardware and software level to address the challenges of limited field of view in AR applications. 

Hardware solutions aim to expand the user's peripheral vision. For example, RadialLight~\cite{gruenefeld_radiallight_2018} utilizes LED plugins to extend the field of view, while projects like Eye-q~\cite{Costanza} introduce wearable peripheral displays embedded in eyeglasses. 

On the software side, visual notifications have been employed to mitigate the limitations of limited field of view. Projects such as the Flying Arrow~\cite{gruenefeld_flyingarrow_2018} utilize a moving arrow indicator to direct the user's attention to out-of-screen objects. Matsuzoe et al.~\cite{Matsuzoe2017} explored the use of vibrating icons to indicate the proximity of out-of-screen targets. Advancing this research further, Qian et al.~\cite{qian_restoring_2018} tested different notification approaches in search tasks using AR glasses and found that edge notifications with a blue indicator yielded better performance. 

In addition to notifications, projects like Push the Red Button~\cite{Plabst2022} have shown the effectiveness of wrist-based notifications when combined with other AR elements. Notifications displayed within the user's field of view are quickly noticed and understood if the user's view direction is known. Meanwhile, the introduction of a mobile peripheral vision model, as studied in The Peripheral Vision study~\cite{Chaturvedi2019}, aims to simplify multitasking on smart glasses without requiring additional hardware. Rzayev et al.'s research~\cite{Rzayev2020} focuses on displaying notifications during face-to-face communication, which further highlights the importance of tailoring notification strategies for different scenarios within AR applications.

While previous research has primarily focused on navigation tasks and object finding, our study specifically targets how users perceive avatar behavior in musical activities. Drawing from previous notification research we measure reaction time, co-presence~\cite{piumsomboon_mini-me_2018,biocca2002defining}, and enjoyment as measurements of effective notification strategies for musical tasks in XR. Leveraging software notifications with AR glasses, we aim to address the limitations inherent to AR glasses and enhance the musical collaboration experience. 

\begin{figure*}[!ht]
 \centering 
 \includegraphics[width=\linewidth]{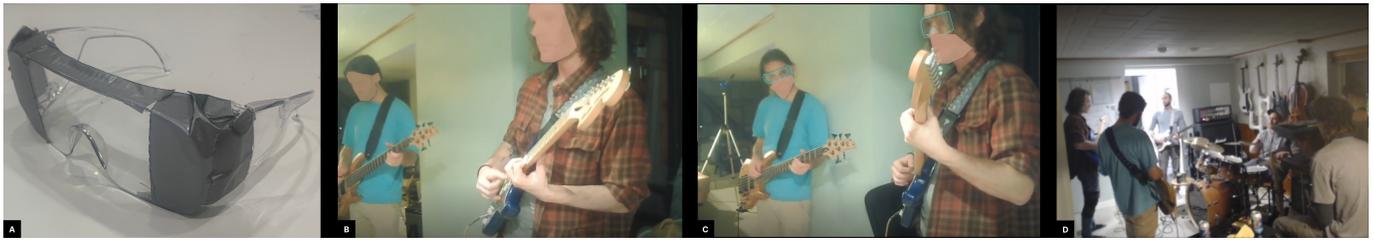}
 \caption{(A) limited FOV Goggles  (B) Normal Jam Session:Musicians with full FOV   (C)limited FOV:Musicians Jamming with the Goggles (D)Jam Session }
 \label{fig:JamSession}
\end{figure*}

\subsection{Social presence with the virtual Avatars in AR}

The importance of interacting with avatars in AR research is highlighted, and understanding users' perception of avatar presence is crucial for assessing its impact in musical collaboration \cite{gorlich2022societal}. Previous studies have utilized social presence questionnaires to evaluate the sense of presence \cite{piumsomboon_mini-me_2018,Hopkins22,biocca2002defining,hwang_incorporating_nodate}. Osmers et al. \cite{Osmers2021} conducted a review study supporting the influence of social presence in asymmetric cooperation settings on user preferences and acceptance.

Furthermore, Hart et al. \cite{Hart2021} found a preference for 3D avatars in XR communication scenarios, leading us to focus on 3D avatar musicians in our study. In an AR drum circle, participants paid a lot of attention to the avatar's body expression, as observed by Hopkins et al. \cite{Hopkins22}. These findings inspired us to explore software modifications for AR such that the avatar's full body appears on the external screen.

Similarly, eye contact and body gestures have been shown to improve task performance in several studies \cite{KurzwegWolf+2022+175+201,Bai2020,kimmel23,Greenwald2017,Blattgerste2017}. However, the limited field of view (FoV) in AR diminishes avatar social presence and the user's perception of the avatar's message \cite{Shin2022}. Shin et al. \cite{Shin2022} found that users wearing AR head-mounted displays (HMDs) faced challenges due to frequent head movements caused by limited FOV. Piumsomboon et al.'s Mini Me project \cite{piumsomboon_mini-me_2018}, which compared social presence with AR glasses between a human-size avatar and a mini avatar on a screen in various scenarios, showed that Mini Me achieved a higher social presence score. Their research scenarios included urban planning, design tasks \cite{Bai2020}, meeting rooms \cite{KurzwegWolf+2022+175+201}, and tea parties \cite{piumsomboon_mini-me_2018}.

In our research, we specifically focus on musical activities with multiple avatars that are outside of the FOV. Building upon \textit{prior research}, \textit{observations of a musical jam session}, and \textit{player feedback}, we created a \textbf{Mini Musicians Avatar} that appears when the life-sized avatar is outside of the player's FOV.



\section{Methods}
In our study we investigate how FOV affects musical collaboration. First, we observe a live music jamming session with and without glasses that restrict the FOV to 52$^\circ$. We recorded the whole session and analyzed the video. Post-analysis and interview, we built the experiment apparatus according to the analyzed result from the jam session. This method of observation was inspired by musician video observation methods conducted by Hopkins, et al \cite{Hopkins22}.

\subsection{ Observation of a Jamming Session}
\label{section:MT}
To examine the impact of limited field of view (FOV), we conducted two 30-minute jam sessions with two guitar players, a bass player, a drummer, and a singer (Figure \ref{fig:JamSession} D). The first session was a control with no external factors (Figure \ref{fig:JamSession} B). In the second session, participants wore goggles restricting their vision to 52$^\circ$, simulating AR goggles (Figure \ref{fig:JamSession} C). Both sessions were recorded with three web cameras from different angles. We assessed music quality, performance coherence, and logged participants' body positioning and language. Two minutes of each session were analyzed per participant by a research assistant using Datavyu and written notes \cite{datavyu}.

\subsubsection{Musicians}
The jam session group consisted of five musicians aged 25 to 31: two guitarists, a bassist, a keyboardist, and a drummer. Each had over five years of experience with their instruments and regularly jammed together once or twice a week.

\subsubsection{Normal jamming}
During a regular jamming session, the musical coordination was excellent, with a high level of synchronization and a rapid tempo. Although there were a few minor miscommunications, they were resolved swiftly. The participants engaged in frequent communication, often making quick eye contact and occasionally exchanging minimal verbal cues. Throughout the session, each musician maintained multiple instances of eye contact with other musicians and their respective instruments.

\subsubsection{Limited FOV jamming}
In the jam session with a limited FOV, the musical collaboration was observably worse compared to the previous session. The music played had reduced coherence and experimentation, as all the musicians adopted a more conservative approach. Communication during this session was also hindered, with fewer instances of actual eye contact among the musicians. In fact, some musicians stopped looking up altogether. Additionally, there was an increase in overall body movement, including frequent head turning. For instance, the drummer had to move their head extensively to see the drums properly, affecting their playing.

\subsubsection{Summary of Observation}
Between the two, musical coherence was better in the control test over the test with the vision-restricting goggles on. The body language was also more open (head up, bodies turned toward each other, etc) in the first test. Participants looked down a lot more with the goggles on and made less eye contact. It was also observed that the drummer struggled to see their entire drum kit with the goggles on as the kit was wider than their FOV. 

The number of times participants either looked up or around at other participants were further analyzed. This was defined as clearly looking over to another participant's instrument or eyes. For this test, 2 minutes of each session were watched, observing one participant. Then, it was repeated for the same 2 minutes of the session for the other participants. For the control session, the participants had a total of \textbf{62} times when they looked at another participant. For the other session, the participants had a total of \textbf{27} times when they looked at another participant. This shows that the limited FOV caused less non-verbal communication, due to a lack of peripheral vision.



\subsubsection{Summary of Post-Jam Session Interview}
We conducted an analysis of interviews with the participants regarding their experiences of jamming with limited FOV compared to normal jam sessions. The main takeaways from the interviews are as follows:

\textbf{Control Jam:} Participants who experienced a normal jamming session reported that the jam went well and felt sufficiently representative of their usual jams. 

\textbf{Limited FOV Jam:} Initially, participants who jammed with limited FOV did not feel much difference compared to the control jam. However, they gradually started noticing certain limitations. Specifically, they mentioned that they found themselves focusing straight ahead and not looking around. This narrowed focus caused some communication breakdowns as players would often take longer to notice that someone was looking at them or trying to convey something.

\textbf{Missing Visual Cues:} One participant expressed frustration, stating that they don't know what they are missing because they cannot see it due to the limited FOV. This inability to perceive the entire visual scene hindered their overall experience and understanding of the jamming session.

\textbf{Forgetting Presence:} Another noteworthy observation was that participants mentioned how, with the limited FOV, they sometimes forgot that certain players were present in the virtual space. This forgetfulness could potentially lead to unintentional exclusion or overlooking the contributions of certain individuals during the jam.

\textbf{Desire for Mini Avatars:} Three participants mentioned that they would love to see a mini version of the avatars, similar to those in Mario Kart. They suggested that having smaller avatars displayed in a corner of their field of view could help them maintain awareness of other players and enhance communication during the jam.

Overall, the interviews revealed that while the limited FOV initially did not feel drastically different, participants gradually became aware of the constraints it imposed. The reduced ability to observe the entire scene and the potential for communication breakdowns were notable drawbacks. Participants expressed a desire for alternative visual cues, such as mini avatars, to enhance their awareness and communication in the limited FOV setting. These insights provide valuable feedback for the design and improvement of collaborative AR experiences, particularly in the context of musical activities. We decided to build and test a mini avatar system, as it was consistent with previous work \cite{piumsomboon_mini-me_2018} and desired by several interviewed musicians.

\subsection{Limitations of Prior Work}

From previous work, Weng et al. \cite{Weng_HolojamVR_2023} conducted an experiment testing the limitations of FOV. Their results indicated that the holographic display without FOV limitations achieved the highest co-presence score, the lowest response time, and the highest task enjoyment score. This suggests that FOV limitations negatively affect musical collaboration with avatars. Additionally, while AR glasses with notifications may improve co-presence, they could potentially reduce enjoyment. In subsequent work, Weng et al. \cite{Weng_HolojamISMAR_2023} enhanced the notifications by introducing "mini musicians."In our paper, we present more details on the system implementation and additional results from our user study.

\begin{figure*}[!ht]
 \centering
 \includegraphics[width=1\linewidth]{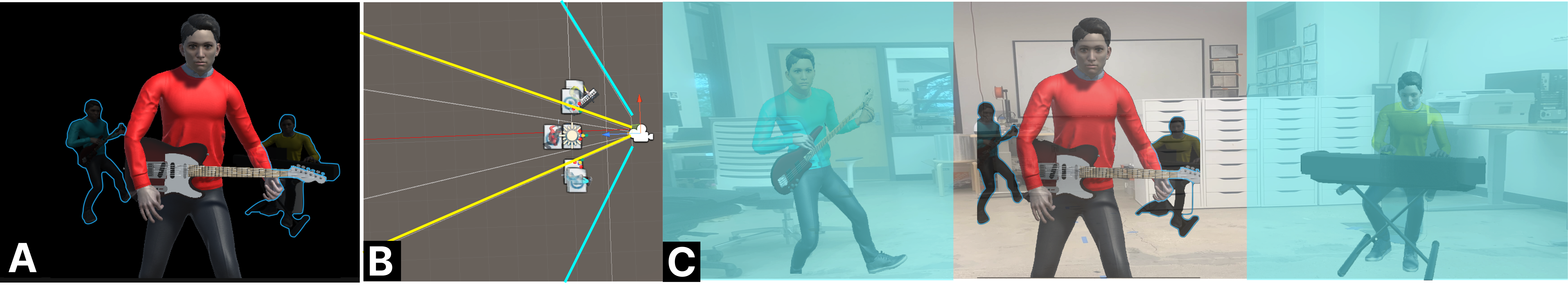}
 \caption{(A) Virtual Camera views
(B) Virtual Cameras setup
Yellow lines: The range of the limited FOV
Blue lines: A wider range of FOV
(C) Mini musicians: AR glasses view
}
 \label{fig:MiniMusicians}
\end{figure*}

\section{Research Questions}
The key insights from prior studies are summarized below:
\\
\textbf{XR technology has transformed traditional musical interactions}, allowing musicians and audiences to engage with virtual objects and environments. Mixed reality technologies were explored as a means to create a sense of presence for musicians. \\
\textbf{Social presence with virtual avatars in AR is crucial for assessing their impact on musical collaborations}. Previous research has shown the significance of social presence in asymmetric cooperation settings, and avatars in 3D form have been preferred in XR communication scenarios. \\
\textbf{Opportunity to study musical activities with multiple avatars}, examining the impact of avatar behavior on user perception and exploring software notifications with AR glasses to address their limitations.

Building upon these findings, our research questions are:

   \begin{itemize}
       \item RQ1: How fast can a player notice an avatar musicsian when the avatar is communicating with the player non-verbally (avatar is looking up at the player) with and without a limited FOV?
       \item RQ2: What are the Aggregated Social Presence Scores (given by Aggregated Social Presence Scores questionnaire \cite{biocca2002defining}) with the avatars with and without a limited FOV?\cite{Weng_HolojamISMAR_2023}
       \item RQ3: Does a limited FOV affect task enjoyment?
   \end{itemize}

\section{Studying the Effects of Notifications in Limited FOV Musical Scenarios }
To answer our research questions, we redesigned the previous pilot study and conducted a user study using our improved system.

\subsection{System design: HoloJam and Mini Musicians} 

According to the interview and previous research we built two XR hardware setups enabling players to jam with virtual avatars. As successfully implemented in pilot studies, the first system was created without a limited FOV using holograms. The second system used AR glasses running an application with a limited FOV of 52$^\circ$. Using AR glasses we had two conditions: one with mini musicians (Figure \ref{fig:teaser} A)and one without any additional content (Figure \ref{fig:teaser} B). 

\subsection{HoloJam}

The first setup was a Hologram Jam (Figure \ref{fig:teaser}
 C), utilizing a Pepper's ghost illusion. We chose this setup over a screen display because previous research indicates that the Pepper's ghost illusion provides a more immersive experience, particularly in a musical performance environment \cite{Trajkova_HoloDance_23,Lee_AvatarConcet_23,Hamanaka_MelodySlotMachine_19}. Furthermore, according to the works of XR Jam \cite{Hopkins_XRJAM_2024} and Schlagowski et al. \cite{Schlagowski2023}, musicians tend to move around while playing. The hologram setup offers two important features: it gives users a sense of real space and allows in-person players to perform alongside avatars without being occluded by a screen.(e.g. stands behind the avatars)
The hardware setup consisted of a laptop PC equipped with an Intel 10700k CPU, an Nvidia 1070 GPU, and 32GB of memory. We used a transparent acrylic board placed at a 45-degree angle from the screen to reflect the television’s visuals as a hologram screen. This setup enabled participants to interact with the avatars without the limitations of a restricted FOV.

 \subsection{AR Glasses}
The second was a jam session using NReal AR glasses (Figure \ref{fig:teaser} E) \footnote{https://www.nreal.ai/light/}. In the AR glasses set up, we placed three virtual avatars in the AR environment. One in the middle, and the other two avatars on both sides beyond 52$^\circ$ (outside FOV). 



\subsubsection{Notification Design Mini Musicians}
The notification method we used was the Mini Musicians. This design was adapted from the Mini Me research \cite{piumsomboon_mini-me_2018}, which we modified for a musical context to enhance the user experience in musical collaborations.

In our virtual environment setup, we implemented a second virtual camera with a wider FOV than the main camera (Figure \ref{fig:MiniMusicians}
 B). This second camera was placed on a 2D transparent canvas positioned in front of the main camera. The main camera displayed the primary view (Figure \ref{fig:MiniMusicians} A), while the second camera provided an extended view, showing smaller objects (in this case, the avatars) that were outside the FOV of the main camera as the user moved their head to look around (Figure \ref{fig:MiniMusicians} C).

This camera setup offers for AR developers. By incorporating a second virtual camera, developers can adapt the off screen notification system to various contexts and content without needing to redesign the entire notification method. This approach ensures that users maintain a wider vision in the AR view, enhancing their ability to interact with and respond to virtual elements during musical collaboration.Additionally, this method provides a more immersive experience by maintaining awareness of avatars outside the immediate FOV, thereby improving the overall effectiveness of AR notifications.

\subsection{Interaction and Content Design}

\subsubsection{Interactive Device for the Participants}
In this apparatus, we provided two input devices for the participants. The first one was a MIDI drum pad that played the drum sound from a speaker . The second one was a foot pedal (Figures \ref{fig:teaser} E and F) that indicated three positions of the avatar which let the user select by step on the cross response position of the pedal, which avatar was following them.


\subsubsection{XR Content design}
We designed the musical collaboration scenarios that enabled participants to play musical instruments with virtual avatars.It was a jam session with three animated avatars playing different instruments: bass, guitar and Keyboard (Figure \ref{fig:teaser} D). The animations reacted to the background music. In both scenarios the avatar models and several animations were downloaded from Mixamo \footnote{https://www.mixamo.com/}. The reactive animations were controlled by an animated rig plugin in the Unity Game Engine\footnote{https://unity.com/}.


\section{User Evaluation}
According to the pilot study results we improved our AR notification system by solely including mini musicians, rather than light or symbological indicators. Furthermore, we simplified the experiment procedure to reflect a more thorough design approach and mechanism of data collection.

\subsection{Experimental Design}
To address our research questions we conducted an experiment that compared musical experience and social presence with avatars in different FOV situations in AR. We conducted a 2x3 within-subjects experiment to simulate a jam session with three avatars. In the AR conditions, we used two contexts: 1) visual notifications use mini musicians to indicate avatar gesture when outside of the FOV. The avatars have a single gesture, to look up at the participant from the default position of looking at the avatar's own instrument, and 2) no visual notification. In the holographic setup, we put three holograms in front of a participant with the same FOV as the AR glasses conditions. Like previous experiments the hologram condition simulates playing the Jam session with avatars, but without an FOV limitation.
We let participants play a midi drum because the drum playing is easier to a regular participant than other melody instruments.

\subsection{Participant}
  
A total of 19 participants were recruited for the study through email lists and word of mouth. All participants were university students, and the experiment took place in a research laboratory on the university campus. Each participant got 5 dollars cash compensation for their time.
Among the participants, 11 were undergoing professional music training. The number of training years varied widely, ranging from 0 to over 10 years. The majority of participants had 2-3 years of training. 

\subsection{Procedure}
The experiment ensured informed consent and explained the drumming task, equipment, and data handling. Participants synchronized drumming with music while identifying an avatar looking at them (pressing a pedal button).

Three conditions were tested: hologram (participants adjusted their position in front of 3 holograms with lights off), and two AR glasses setups (researchers helped participants wear glasses and adjusted the environment).

A counterbalanced Latin square design ensured each participant experienced all conditions in a random order (each lasting 8 minutes). Questionnaires assessed social presence, enjoyment, and post-test feedback. The entire session averaged 45-50 minutes.

\section{Results}

We measured three major quantitative variables. The first was reaction time, measured as the amount of time it took the participants to notice an avatar was looking at them. The second was enjoyment, which employed a 1-7 likert scale to measure whether the participants were enjoying the tasks or not. The last quantitative measure obtained was the social presence aggregate score. Finally, we collected the qualitative data from the participants' free writing of at least 150 words \cite{Hopkins22,Slater2023}.
\begin{figure*}[!ht]
 \centering
  \includegraphics[width=1\linewidth]{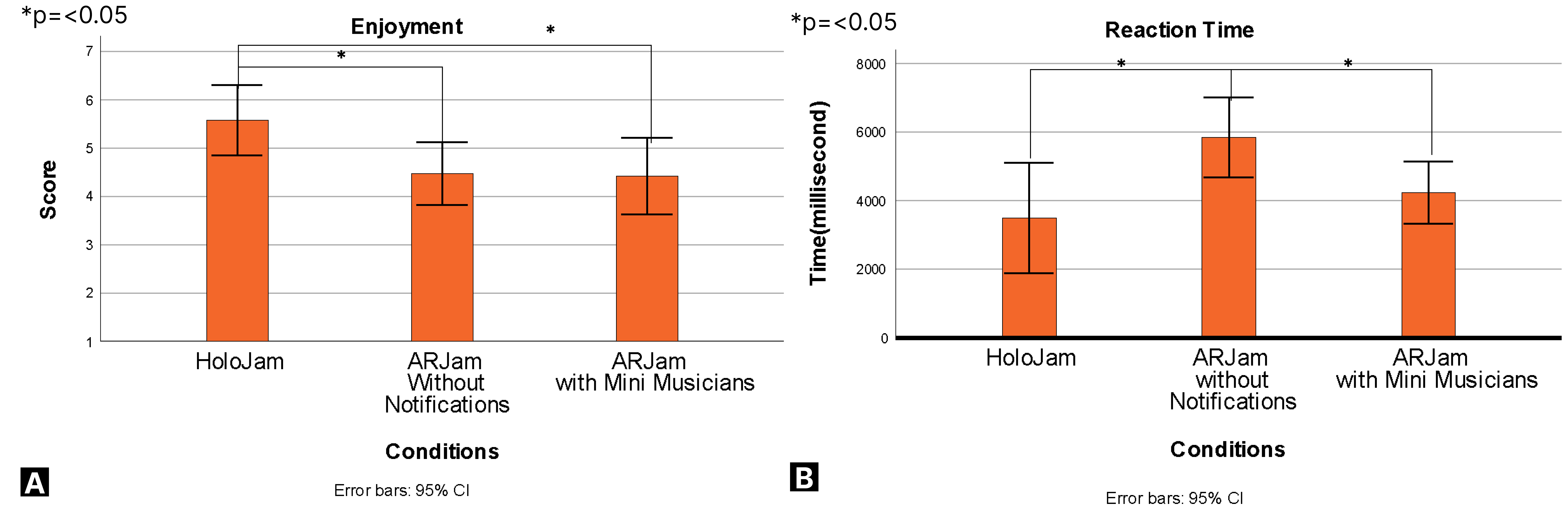}
 \caption{ (A)Enjoyment:Enjoyment score (B)Reaction Time: the time a participant took to notice an avatar was attempting to communicate}
 \label{fig:reactionandEnjoyment}
\end{figure*}

\subsection{Reaction Time}

To answer our \textit{RQ1},we measured the response time when the participants noticed the following avatar. We collected 10 rounds of data from each participant and measured the mean response time in milliseconds after all rounds. To analyze the data, a repeated measures Analysis of Variance (RM-ANOVA) was conducted on the data. The result showed that mean enjoyment differed significantly across three conditions F(2, 36) = 4.476, p = 0.018) (Figure \ref{fig:reactionandEnjoyment} B). The hologram condition had significantly faster reaction times than the AR glasses condition without notifications ($\eta^2_p$=0.34). As well as AR glasses with mini musicians was also significantly faster than the no notification condition ($\eta^2_p$=0.20).
Based on the results, we can say that the limited FOV does affect the time it takes for the user to notice the avatar (\textbf{RQ1}).Additionally, the results show that the Mini Musicians helped participants notice the avatar faster than in the no-notification condition.

 \subsection{Social Presence Aggregate Score}
In the previous works Weng et al.\cite{Weng_HolojamISMAR_2023} address \textit{RQ2}, researchers assessed the Social Presence Aggregate Score, finding significant differences in all categories (Co-Presence, PMU, PBI) across conditions, particularly favoring the hologram condition. The hologram condition consistently scored higher, indicating that a limited FOV impacts the social presence in a musical collaboration. No significant differences were noted between Mini Musicians and No Notifications.

\subsection{Enjoyment}
After the participants finished each task, we gave a task enjoyment questionnaire. Quantitative scores between 1 and 7 (1 - lowest, 4 - neutral, and 7 - highest) were indicated for measure enjoyment. The result showed that mean reaction time differed significantly across three conditions (F(2, 36) = 13.698, $p \leq 0.001$) (Figure \ref{fig:reactionandEnjoyment} A). The hologram condition had significantly better enjoyment scores than the AR glasses condition without notifications ($\eta^2_p$ $\leq 0.01$) and the AR glasses condition with mini musicians ($\eta^2_p$ $\leq 0.01$). There were no significant differences between Mini Musicians and Without notifications.In this result, users without a limited field of view (FOV) experience better enjoyment with the jamming experience (\textbf{RQ3}).


\subsection{Participant responses }
The analysis of participants' free writing response from the three jamming experiences—AR jamming, AR jamming with Mini Musicians notifications, and Holojam—revealed significant differences in terms of field of view, immersion, comfort, and preference. The AR jamming experience had limitations related to the narrow field of view and discomfort caused by wearing glasses. Despite these limitations, users still found the AR experience immersive and appreciated the detailed rendering of avatars. The introduction of notifications in AR jamming aimed to address the field of view issue but received mixed responses. While some users found the notifications helpful, others found them distracting or ineffective. In contrast, HoloJam offered the most natural and immersive experience, with a broader FOV, better avatar interaction, and increased comfort. Users preferred the HoloJam experience over the AR versions due to its convenience and realism.

The implementation of notifications in the AR application aimed to improve the limited field of view and bridge the gap with the holographic experience. However, the feedback indicated that the notifications did not significantly enhance the overall user experience. Users found them confusing, distracting, or lacking impact. Some users reported losing focus on the main avatars and split their attention between the primary and notification fields. The holographic experience remained superior in terms of immersion, natural feel, and the ability to monitor all avatars simultaneously. While the notifications did have positive aspects such as assisting with avatar focus, providing directional guidance, and adding awareness, they fell short of fully bridging the gap between AR and holographic experiences.

\section{Discussion}
Based on our findings, we discuss several key topics that require further exploration and in-depth analysis to better understand their implications

\subsection{Improved Musical Collaboration with holograms}
Our results clearly demonstrate the significant impact of the limited Field of View (FOV) on the quality of musical collaboration within virtual avatar-based XR experiences. The data we have gathered strongly suggests that holographic representations of avatars offer substantial advantages, including heightened enjoyment, enhanced co-presence, and swifter reaction times during collaborative musical endeavors.
Drawing upon our extensive field observations and research findings, it becomes evident that effective communication plays a pivotal role in enhancing musical performance in XR environments. Musicians who experienced more seamless and immersive communication with holographic avatars consistently exhibited superior musical performance. This finding underscores the importance of addressing communication challenges within XR setups to optimize collaborative experiences.
Moreover, the feedback from our participants is in line with the quantitative data, as they reported a noticeable ease and comfort while engaging with holograms compared to the alternative of interacting with virtual musicians and their own physical instruments. This sentiment further bolsters our argument that holograms represent a superior choice for facilitating musical collaboration with avatars within the XR landscape, ultimately outclassing the capabilities of AR glasses in this particular context.

\subsection{AR Glasses Notifications in Musical Collaboration}

In previous research, Weng et al. \cite{Weng_HolojamVR_2023} experimented with abstract notifications, specifically the blue beam notification. However, the results showed that the abstract notification did not significantly affect the drum circle setup in terms of co-presence, enjoyment, or reaction time. This outcome is attributed to the abstract notification's failure to provide relevant information about avatars. Subsequently, we introduced mini musicians, a notification depicting miniature avatars, which improved reaction time compared to no notification. Despite this improvement, neither notification enhanced the sense of co-presence and enjoyment. Participants reported that the notifications caused distraction and confusion. Consequently, we suggest that further improvements are necessary when using notifications for musical collaboration in AR glasses

\subsection{Exploring the Potential of Social Cue Notifications for AR Glasses}

In this study, we focused on investigating the effectiveness of out-of-field-of-view avatar notifications for AR glasses in jamming sessions. Our research specifically targeted social co-presence and enjoyment, rather than functional designs like navigation, task alarms, or instructions in AR glasses. The limitations imposed by FOV continue to be a challenge for state-of-the-art head-mounted XR devices. How can we inform users about out-of-FOV avatars who wish to communicate with them? We propose that there are numerous research questions that can be explored in this specific field to address this issue.

\section{Limitations and Future Work}
This study had several limitations that we aim to address in future research to further enhance our understanding and expand the possibilities of musical collaboration in XR.

\subsection{ Musical Collaborations with XR devices}
Future studies could focus on exploring social notifications specifically designed for AR glasses. By integrating notifications tailored to the AR glasses platform, researchers can enhance the overall user experience and investigate the impact of social cues on musical collaboration. While our study primarily concentrated on see-through AR glasses, there are other Mixed Reality (MR) devices available that employ video pass-through technology, offering a wider field of view. Future research should consider examining the effects of these devices and compare them to see-through AR glasses to gain a comprehensive understanding of their impact on musical collaboration.

\subsection{Intelligent Avatar Interaction}

Previous research by XR Jam \cite{Hopkins_XRJAM_2024} introduced a musical collaboration system that integrates Artificial Intelligence (AI) avatars. This innovative approach allows AI avatars to participate in musical sessions, responding to the users' musical inputs. Future research could delve deeper into the potential of AI avatars to interact with the system. This involves investigating how intelligent avatars can dynamically respond to users' music playing, offering real-time feedback and interaction. The integration of AI could lead to more fluid and natural interactions, fostering a richer and more immersive environment for musical collaboration.

\subsection{Social Scenarios for XR Explored}
Our study primarily focused on the musical jamming experience, but there are other scenarios in which AR glasses could be utilized. Future work should expand its scope to explore the development of notifications for general situations that AR glasses users may encounter. This would provide a more comprehensive understanding of the potential applications and benefits of AR glasses beyond musical collaboration.

\subsection{Evaluation Methods}
While self-reported questionnaires provided valuable insights into participants' subjective experiences, future research should consider incorporating more objective data. For example, measuring participants' skin condition or utilizing brainwave monitoring devices can offer additional quantitative and physiological insights into their level of engagement, immersion, and cognitive load during the collaborative musical experience.

\subsection{Future Study}
A follow up study we would like to conduct is one that involves eye-tracking during live jam sessions. High-density eye-tracking has the potential to reveal a lot of information about the dynamics of eye-gaze during communicative tasks in AR and in live jam sessions. A comparative analysis of eye-gaze between the conditions and a normal jam session may give us in-depth insight into what is missing from the AR experience and how we can better overcome these challenges. 

Additionally, interview results suggested that players experienced communication breakdowns when field of view was occluded, suggesting that visual communication is being used to recognize and correct musical mishaps. The mechanism of collective correction of the music will be studied in future experiments.

\section{Conclusion}
The objective of our study was to investigate the impact of field of view (FOV) on players' perception of communicative gestures from their partner's avatar in a musical task. Additionally, we designed various notifications for AR glasses. Our system setup involved the creation of HoloJam, an FOV-unlimited holographic installation, along with AR glasses applications featuring Mini Musicians notifications. Through our conducted study, we aimed to address our research questions. The results of our study indicated that HoloJam outperformed other setups across all measures. Additionally, the Mini Musicians notification for AR glasses contributed to reduced reaction times. Therefore, we conclude that FOV does indeed influence musical collaboration with avatars in extended reality (XR) environments. Furthermore, the use of notifications assists users in perceiving avatars more quickly. Nonetheless, there remains ample room for exploration regarding social interaction with out-of-FOV avatars in XR.



%
%


\section*{Acknowledgment}

The authors would like to thank Gregoire Phillips, Alvin Jude, Gunilla
Berndtsson, Per-Erik Brodin, Amir Gomroki, and Per Karlsson for
their support and advice throughout this project. This project was supported by a grant from Ericsson Research, and by the National Science Foundation under Grant No. IIS-2040489.


\bibliographystyle{IEEEtran}
\bibliography{paper}

\end{document}